# Approximation by Quantum Circuits

E. Knill, knill@lanl.gov[*]

May 1995

## Abstract

In a recent preprint by Deutsch et al. [5] the authors suggest the possibility of polynomial approximability of arbitrary unitary operations on $n$ qubits by 2-qubit unitary operations. We address that comment by proving strong lower bounds on the approximation capabilities of $g$-qubit unitary operations for fixed $g$. We consider approximation of unitary operations on subspaces as well as approximation of states and of density matrices by quantum circuits in several natural metrics. The ability of quantum circuits to probabilistically solve decision problem and guess checkable functions is discussed. We also address exact unitary representation by reducing the upper bound by a factor of $n^2$ and by formalizing the argument given by Barenco et al. [1] for the lower bound. The overall conclusion is that almost all problems are hard to solve with quantum circuits.

## 1 Introduction

There has recently been great interest in the properties of quantum computation and quantum circuits, partly due to the proof by Shor [9] that quantum computation can be used to efficiently factor and compute discrete logarithms in modular arithmetic. An important problem in quantum computation is to better elucidate the relationship between quantum complexity and classical complexity. It is well known that a quantum computation can be simulated by a

---

classical Turing machine with at most an exponential speedup. In addition to Shor's result, there is an accumulating body of evidence to support the idea that this exponential speedup is necessary. This includes works by Deutsch [6], Bernstein and Vazirani [4] and Simon [10], which show that relative to certain oracles quantum computation is exponentially more efficient then classical computation, even in the probabilistic model.

Here we address some open issues in quantum circuit complexity. In classical circuit theory a simple counting argument shows that almost all functions from $2^{[n]}$ (the space of $n$-bit patterns) to $2^{[n]}$ are hard in the sense that to compute them requires exponentially many gates. In quantum circuit theory, a similar result is known to hold for exact computation. However, the question of an exponential lower bound for approximate or probabilistic computation of both classical and non-classical problems had not been answered. In fact, in a recent paper by Deutsch, Barenco and Ekert [5], an intuitive argument is given for why they believe that approximate computation with quantum circuits might be possible with polynomially many two-bit gates. In this paper we show that this is not the case. In the process we investigate the overall difficulty of various related computation tasks in the quantum circuit model.

We present a general approach to studying worst case quantum circuit complexity by considering the problems of representing and approximating actions of unitary operators on arbitrary sets of orthonormal vectors in the Hilbert space. We consider approximation of states in the usual norm as well as approximation of density matrices in a fairly weak norm. For the former problem we show that unless the number of gates is exponentially large, almost all states are approximated no better than random by any circuit. For the second problem we use the total variation distance and obtain a similar (though suboptimal) result.

We define two types of classical problems, one which requires a probabilistic solution and another which requires a checkable solution. In each case we give explicit bounds on the measure or number of problems that can be solved by quantum circuits with a given number of gates. The bounds all imply that almost all problems are exponentially hard for quantum circuits.

An interesting consequence of this work is that even the approximate state conversion problem is exponentially difficult. That is, for almost all states (and almost all pairs of states and density matrices), the smallest quantum circuit which converts the first state to an approximation of the second requires exponentially many gates. This should be compared to the classical theory of *relative information distance* [3]. It reflects the additional information that is implicitly available in quantum states.

An outline of the paper follows. Section 2 summarizes the notation and quantum computation background required for this work. In Section 3, we revisit the upper and lower bounds for exact unitary representation given in [1], improving the upper bound by a factor of $n^2$ and proving the lower bound in [1]. Section 4 covers approximation of the action of unitary operators on members of a basis in the usual Hilbert space norm. We also obtain bounds for approximation of density matrices from the point of view of measurement. In Section 5 we consider two types of classical problems and count the number of such problems that cannot be solved with bounded size quantum circuits.

## 2 Preliminaries

A basic understanding of Hilbert spaces and measure theory is assumed. The $k \times k$ identity matrix is denoted by $I_k$ or simply $I$ if $k$ is understood. A diagonal matrix with diagonal elements given by $\mathbf{d} = \langle d_1, \ldots, d_k \rangle$ is denoted by $D(\mathbf{d})$.

**Quantum circuits.** To understand the problems in quantum circuit complexity requires some knowledge of quantum circuits and their behavior. Descriptions of quantum circuits can be found in [12] and [1]. The input space of quantum circuits operating on $n$ *qubits* is a complex Hilbert space of dimension $2^n$ which is represented as an $n$-fold tensor product $Q^{\otimes n}$ of the two dimensional qubit space $Q$ generated by the orthonormal basis $\{|0\rangle, |1\rangle\}$ (using Dirac notation). For example, $Q^{\otimes 2}$ has the standard orthonormal basis

$$\{|0\rangle|0\rangle, |0\rangle|1\rangle, |1\rangle|0\rangle, |1\rangle|1\rangle\},$$

where juxtaposition of vectors denotes tensor product. The standard orthonormal basis is naturally identified with bit patterns in $2^{[n]}$, and we write $|b\rangle$ for $|b_0\rangle\ldots|b_{n-1}\rangle$, where $b_i$ is the $i$'th bit of bit pattern (or binary number) $b$. The expressions $0^n$ and $1^n$ represent the bit patterns of $n$ 0's and $n$ 1's, respectively. A number $k$ used in the context of a bit pattern denotes it's binary representation (in reverse order). In general, we will use $|u\rangle, |v\rangle, \ldots$ for normalized states and $u, v, \ldots$ to denote arbitrary, not necessarily normalized vectors in Hilbert space.

The functions of quantum computation are unitary operations on the input space. The set of unitary operations on $Q^{\otimes n}$ is denoted by $U(Q^{\otimes n})$ or $U_n$ for short. The output of a quantum circuit is therefore an element of $Q^{\otimes n}$.

Note that unitary operations are invertible, so a quantum computation is always reversible. Reversibility is not a strong restriction from the point of view of complexity theory, as discussed in [2, 8].

The correspondence between quantum circuits and classical (reversible) circuits is established by restricting the input of a quantum circuit to the elements of the standard basis, and by *measuring* the output of the circuit. The measurement usually results in projecting the output state $|u\rangle$ onto one of the basis elements $|b\rangle$. The probability of *observing* $|b\rangle$ as the projected output is given by $\text{Prob}(b \mid |u\rangle) = |\langle u|b\rangle|^2$. Thus the general quantum circuit exhibits probabilistic behavior.

In many cases, the relevant domain of a quantum circuit is restricted to states of the form $|b0^{n-l}\rangle$ and instead of measuring all output bits, only the first $l$ bits may be observed. If the output state is $|u\rangle$, the probability of observing that the first $l$ bits are $b$ is $\text{Prob}(b \mid |u\rangle) = \sum_{b'} |\langle u|bb'\rangle|^2$.

**Quantum gates.** The notion of quantum circuit complexity depends on the circuit elements or *quantum gates* that can be used to build a unitary operation. The most natural definition requires that each quantum gate act *locally* on the input space. This means that the unitary operation corresponding to the gate should involve only a small number of qubits. More formally, let $S \subseteq [n] = \{1, \ldots, n\}$. Write $Q_S^{\otimes n}$ and $Q_{\bar{S}}^{\otimes n}$ for the tensor product of the factors of $Q^{\otimes n}$ at positions in $S$ and not in $S$, respectively. Then $Q^{\otimes n}$ is naturally isomorphic to $Q_S^{\otimes n} \otimes Q_{\bar{S}}^{\otimes n}$. We can take a unitary operator in $U(Q_S^{\otimes n}) \simeq U_{|S|}$ and have it act on $Q^{\otimes n}$ by taking its tensor product with the identity. A $g$-qubit gate acting on $Q^{\otimes n}$ is a member of $U_g$ acting on $Q^{\otimes n}$ via a $g$-qubit factor $Q_{\{i_1,\ldots,i_g\}}^{\otimes n}$. We denote the family of $g$-qubit gates by $G_g$. Note that for $g \leq g'$, $G_g \subseteq G_{g'}$.

Quantum circuit complexity, like classical circuit complexity, concerns the question of what operations can be represented by compositions of $b$ gates. The first question to be answered is of course whether the gates chosen are universal. This was solved by DiVincenzo and others [7]: Sufficiently large compositions of members of $G_2$ can represent any unitary operation. Let $G_g^{(b,n)}$ be the set of all unitary operations in $U_n$ which can be represented as a composition of $b$ members of $G_g$. In general, let $\mathcal{V}^{(b,n)}$ be the set of operators in $U_n$ representable by compositions of $b$ operators in $\mathcal{V}$.

**Metrics for $U_n$.** In order to study the complexity of approximating states, unitary operations or other types of functions, we need suitable metrics. There are several choices and in general we will make use of the weakest one. We consider linear operations as matrices over the standard basis. For $A$ a linear operation on $Q^{\otimes n}$, its matrix is defined by $A_{b,b'} = \langle b|A|b' \rangle$. Most of the metrics to be considered are related to the standard Euclidean norm on a Hilbert space $H$ defined by $|u| = \sqrt{u^*u}$.

**The Frobenius norm.** The *Frobenius norm* of $A$ is defined by

$$\begin{aligned} ||A||_F &= \sqrt{\sum_{b,b'} |A_{b,b'}|^2} \\ &= \sqrt{\operatorname{tr} A^* A}. \end{aligned}$$

**The two-norm.** The *two-norm* of $A$ is given by

$$||A||_2 = \sup_{|x\rangle} |A|x\rangle|.$$

**The weak two-norm.** This is a weak version of the two-norm which is useful for quantum computation:

$$||A||_{2,k} = \max_{b<k} |A|b\rangle|.$$

**Total variation distance.** The output of a quantum circuit induces a probability distribution on measurement outcomes. For two probability distributions $\mu$ and $\nu$, the *total variation distance* is defined by

$$v(\mu, \nu) = \sum_b |\mu(b) - \nu(b)|.$$

For two states $|u\rangle$ and $|v\rangle$, their total variation distance on the first $l$ bits is given by

$$\begin{aligned} v_l(|u\rangle, |v\rangle) &= v(\operatorname{Prob}(\cdot \mid |u\rangle), \operatorname{Prob}(\cdot \mid |v\rangle)) \\ &= \sum_{b \in 2^{[l]}} |\operatorname{Prob}(b \mid |u\rangle) - \operatorname{Prob}(b \mid |v\rangle)|. \end{aligned}$$

**Total variation distance for unitary operators.** For unitary operators $U$ and $V$, we can define a total variation distance by

$$v_{l,k}(U, V) = \max_{b<k} v_l(U|b\rangle, V|b\rangle).$$

The following lemmas summarize the relationships between the different distance measures.

**Lemma 2.1** *The following inequalities hold:*

$$\begin{aligned} v_l(|u\rangle, |v\rangle) &\leq v_{l+l'}(|u\rangle, |v\rangle), \\ v_l(|u\rangle, |v\rangle) &\leq 2||u\rangle - |v\rangle|. \end{aligned}$$

**Proof.** We can consider the $v_l$ seminorms as the $L_1$ distance of the probability distributions $P_u(b) = \text{Prob}(b \mid |u\rangle)$ and $P_v(b) = \text{Prob}(b \mid |v\rangle)$ considered as vectors over $b$. Let $P_{x,i}(b) = \text{Prob}(bi \mid |x\rangle)$ for $i = 0, 1$. Then $P_x = P_{x,0} + P_{x,1}$, which implies that

$$\begin{aligned} v_l(|u\rangle, |v\rangle) &= |P_u - P_v|_1 \\ &= |P_{u,0} - P_{v,0} + P_{u,1} - P_{v,1}|_1 \\ &\leq |P_{u,0} - P_{v,0}|_1 + |P_{u,1} - P_{v,1}|_1 \\ &= v_{l+1}. \end{aligned}$$

This gives the first inequality.

To see the other inequality requires going back to the definitions. For a vector $x$ or state $|x\rangle$, let $x_b$ be the projection onto the space spanned by $\{|bb'\rangle\}_{b'}$,

$$x_b = \sum_{b'} \langle bb'|x\rangle |bb'\rangle.$$

Then $\text{Prob}(b \mid |u\rangle) = |x_b|^2$.

$$\begin{aligned} v_l(|u\rangle, |v\rangle) &= \sum_b \left| |u_b|^2 - |v_b|^2 \right| \\ &= \sum_b ||u_b| - |v_b|| (|u_b| + |v_b|) \\ &\leq \sqrt{\sum_b |u_b - v_b|^2} \left( \sqrt{\sum_b |u_b|^2} + \sqrt{\sum_b |v_b|^2} \right) \\ &= 2||u\rangle - |v\rangle|, \end{aligned}$$

where we used the Schwarz inequality in the second to last step. ∎

**Lemma 2.2** *The following inequalities hold:*

$$\begin{aligned} ||A||_{2,k} &\leq ||A||_2 \\ &\leq ||A||_F, \\ ||A||_{2,k} &\leq ||A||_{2,k+k'}, \\ v_{l,k}(U, V) &\leq v_{l+l',k+k'}(U, V), \\ v_{l,k}(U, V) &\leq 2||U - V||_{2,k}. \end{aligned}$$

**Proof.** To see that $||A||_2 \leq ||A||_F$ write

$$\begin{aligned}
|Au|^2 &= \sum_b |\sum_{b'} A_{b,b'} u_{b'}|^2 \\
&\leq \sum_b (\sum_{b'} |A_{b,b'}|^2)|u|^2 \\
&= ||A||_F^2 |u|^2,
\end{aligned}$$

by the Schwarz inequality. The fact that increasing $k$ in the subscript increases the norm is true by definition. The remaining inequalities follow from Lemma 2.1. ∎

According to Lemmas 2.1 and 2.2, if it is difficult to approximate states or operators in a $v_l$ or $v_{l,k}$ (semi)metric, then all the other metrics are difficult too. Note that $v_l(|u\rangle, |v\rangle)$ can be viewed as comparing the density matrices induced by $|u\rangle$ and $|v\rangle$ in the space spanned by the first $l$ bits by comparing only the diagonal of the density matrices.

We will make use of the following lemma which describes the behavior of the 2-norm in conjunction with composition of unitary operators.

**Lemma 2.3** *For $i \in \{1,2\}$, let $U_i$ and $V_i$ be unitary operators. If $||U_1 - V_1||_{2,k} \leq \delta$ and $||U_2 - V_2||_{2,k} \leq \rho$, then $||U_1 U_2 - V_1 V_2||_{2,k} \leq \delta + \rho$.*

**Proof.**

$$\begin{aligned}
||U_1 U_2 - V_1 V_2||_{2,k} &= ||U_1 U_2 - U_1 V_2 - (V_1 V_2 - U_1 V_2)||_{2,k} \\
&\leq ||U_1(U_2 - V_2)||_{2,k} + ||(V_1 - U_1)V_2||_{2,k} \\
&\leq \delta + \rho.
\end{aligned}$$

where unitarity was used in the last step. ∎

**Measures for $U_n$.** To "count" the fraction of unitary operations that can be approximated by $b$-gate circuits requires suitable measures on $U_n$. We will consider two measures, depending on whether we are interested in approximating states or approximating probability distributions. Both measures are induced by measures on sets of $k$ orthonormal states. For a unitary operator $U$, let $\phi_k(U) = \langle\, U|i\rangle | 0 \leq i < k\,\rangle$.

**The sphere measure.** The unit vectors in a Hilbert space $H$ of dimension $N$ are elements of the unit sphere $S_{2N}$ in $2N$ real dimensions. The

Euclidean metric induces a measure $\mu$ on $S_{2N}$ for which

$$\mu(S_{2N}) = 2\frac{\pi^N}{\Gamma(N)}.$$

Let $\mu_1 = \frac{\mu}{\mu(S_{2N})}$ be the normalized measure. Let $\mathcal{U}_k$ be the set of sequences $\mathbf{u} = \langle u_0, \ldots, u_{k-1} \rangle$ of orthonormal states in $H$. One can view an element $\mathbf{u} \in \mathcal{U}_k$ as being obtained by choosing $|u_0\rangle$ on the unit sphere in $Q^n$, then choosing $|u_1\rangle$ on the induced unit sphere in the subspace orthogonal to $|u_0\rangle$ and so on. This induces a measure on $\mathcal{U}_k$ by integrating functions "inside out" with respect to $\mu_1$, that is by integrating over the $k$'th element of the sequence of states first. This measure is denoted by $\mu_k$. For measurable functions $f(\mathbf{u})$,

$$\int d\mu_k(\mathbf{u})f(\mathbf{u}) = \int d\mu_{k-1}(\langle u_0, \ldots, u_{k-2} \rangle) \int d\mu_1(u_{k-1})f(\mathbf{u}),$$

where $\mu_1$ in the inner integral is over a sphere of dimension $2(N-k+1)$. The measure $\mu_k$ induces a measure on $U_n$ via $\phi_k$. We denote this measure by $\mu_k$ as well. Note that the induced measure on $U_n$ is defined on a $\sigma$-subalgebra of the Borel measurable sets. For our purposes, we will extend $\mu_k$ to all subsets of $U_n$ by defining $\mu_k(X) = \bar{\mu}_k(\phi_k(X))$, where $\bar{\mu}_k$ is the outer measure.

For us, the most important property of $\mu_1$ is the measure of the sphere within $\epsilon$ of a state. For any vector $u$, let $B_{u,\epsilon}$ be the set of vectors $v$ satisfying $|u - v| \leq \epsilon$.

**Lemma 2.4** *Let $|u| = 1$ and let $S$ be a unit sphere in a subspace of complex dimension $m$. Then for $\epsilon < \sqrt{2}$ and $m \geq 3$,*

$$\mu_1(B_{u,\epsilon} \cap S) \leq \frac{\left(\epsilon\sqrt{1 - \epsilon^2/4}\right)^{2m-1}}{\sqrt{2m-1}(1 - \epsilon^2/2)}.$$

The proof of Lemma 2.4 is in the appendix.

**The density measure.** Consider the map $\text{Prob}_l$ which takes a vector $u$ to the vector $\text{Prob}_l(u)$ defined by $\text{Prob}_l(u)_b = \sum_{b'} |u_{bb'}|^2$ where $b$ is an $l$-bit string. For unit vectors $u$, $\text{Prob}(u)$ is in the simplex $\Delta(2^l)$ of dimension $2^l$ defined by $\sum_b \text{Prob}(u)_b = 1$ and for each $b$, $\text{Prob}(u)_b \geq 0$. The Euclidean metric induces a measure $\nu$ on $\Delta(N)$ for which $\nu(\Delta) = \frac{\sqrt{N}}{\Gamma(N)}$. Let $\nu_{l,1} = \frac{\nu}{\nu(\Delta)}$ be the normalized measure. This measure induces a measure $\nu_{l,k}$ on $\mathcal{U}_k$ via the product measure and the map $\mathbf{u} \mapsto \langle \text{Prob}_l(u_0), \ldots, \text{Prob}_l(u_{k-1}) \rangle$. Note that if $2^n \geq k2^l$, then this map is onto so that the induced measure is non-trivial. This again induces a measure on $U_n$, also denoted by $\nu_{l,k}$, via $\phi_k$. This measure is extended to arbitrary subsets as we did for $\mu_k$.

We need an analog of Lemma 2.4 for $\nu_{l,1}$. For any state $|u\rangle$, let $B_{u,l,\epsilon}$ be the set of states $|v\rangle$ satisfying $v_l(|u\rangle, |v\rangle) \leq \epsilon$.

**Lemma 2.5** $\nu_{l,1}(B_{u,l,\epsilon}) \leq (2\epsilon)^{2^l - 1}$.

The proof of Lemma 2.5 is in the appendix. The result can be strengthened for $\epsilon \geq \frac{1}{2}$, but this is not included in this work.

## 3 Exact representation of unitary operators

DiVincenzo [7] showed that $G_2^{(b,n)} = U_n$ for $b$ sufficiently large. The open problem is to determine

$$b(2,n) = \min\{b \mid G_2^{(b,n)} = U_n\}.$$

Barenco et al.'s [1] analysis shows that $b(2,n) = O(n^3 4^n)$ for DiVincenzo's construction. We will show that $b(2,n) = O(n 4^n)$. They also give an intuitive argument for $b(2,n) \geq 4^n/9 - 1/3n - 1/9$, which we formalize below.

It is interesting to consider a more detailed study of unitary representation by gates. For sequences of orthonormal states $\mathbf{u} = \langle\, |u_0\rangle, \ldots, |u_{k-1}\rangle\, \rangle$ and $\mathbf{v} = \langle\, |v_0\rangle, \ldots, |v_{k-1}\rangle\, \rangle$ let $b(g, n; \mathbf{u}; \mathbf{v})$ be the smallest $b$ such that there exists $U \in G_g^{(b,n)}$ with $U|u_i\rangle = |v_i\rangle$ for each $i$. If the $|u_i\rangle$ are the first $i$ standard basis vectors, then we will write $b(g, n; \mathbf{v})$ for $b(g, n; \mathbf{u}; \mathbf{v})$. Note that if $U \in G_g^{(b,n)}$, then $U^{-1} \in G_g(b, n)$, which implies that

$$b(g, n; \mathbf{u}; \mathbf{v}) \leq b(g, n; \mathbf{v}) + b(g, n; \mathbf{u}).$$

Let $b(g, n; k) = \max\{b(g, n; \mathbf{u})\}$, where the maximum is over all possible choices of the $|u_i\rangle$.

Let $I_w$ be the unitary operation defined by $I_w|0\rangle = e^{iw}|0\rangle$ and $I_w|b\rangle = |b\rangle$ for $b \neq 0$. Another complexity parameter of interest is the following:

$$b(g, n; I_*) = \min\{b \mid \forall w\, I_w \in G_g^{(b,n)}\}.$$

$I_w$ is a special case of the $\wedge_{n-1}(U)$ gates described in [1], where it is shown that $b(2, n; I_*) = O(n^2)$. The reason for considering $b(g, n; I_*)$ is explained by the next theorem.

**Theorem 3.1** $b(g, n; k) \leq k(2b(g, n; 1) + b(g, n; I_*))$.

**Proof.** Let $\mathbf{u} = \langle\, |u_0\rangle, \ldots, |u_{k-1}\rangle\, \rangle$ be $k$ orthonormal states. The proof is in two steps. The first step is to show that the transformation which maps $|i\rangle$ to $|u_i\rangle$ can be extended to a unitary operator with at most $k$ non-identity eigenvalues. The second is to observe that by diagonalizing the unitary operation and exploiting 1-transitivity, it can be decomposed into a product of $I_w$'s, conjugated by operators instantiating 1-transitivity for the eigenvectors.

The first step is accomplished by the following lemma.

**Lemma 3.2** *Let $V$ be a $k \times m$ matrix whose rows are orthonormal. Then there exists an $(m-k) \times m$ matrix $Z$ with orthonormal rows such that the unitary matrix obtained by extending $V$ by $Z$ has at least $m - k$ eigenvalues with value 1.*

**Proof.** Let $W$ be an arbitrary orthonormal basis of the subspace orthogonal to the row space of $V$. The desired matrix $Z$ will be of the form $UW$ for $U$ unitary. The problem is solved if we can determine $U$ and and an $m \times (m-k)$ matrix $X$ with column space of dimension $(m-k)$ such that

$$\begin{bmatrix} V \\ UW \end{bmatrix} X = X.$$

To see that there exists such an $X$ satisfying that $VX$ consists of the first $k$ rows of $X$, it suffices to observe that $V - [I_k, 0]$ has a null space of dimension at least $m - k$. For this $X$, write

$$X = \begin{bmatrix} X_1 \\ X_2 \end{bmatrix},$$

where $X_1$ is $k \times (m-k)$.

We have the following identities:

$$\begin{aligned} X^*X &= X_1^*X_1 + X_2^*X_2 \\ &= I_{m-k}, \\ X_1^*X_1 &= X^*V^*VX, \\ X^*W^*WX &= X^*(I_m - V^*V)X \\ &= I_{m-k} - X_1^*X_1 \\ &= X_2^*X_2. \end{aligned}$$

The identity $(WX)^*(WX) = X_2^*X_2$ implies that there exists a unitary $U$ such that $UWX = X_2$, as desired (see Lemma A.1 in the appendix). ∎

Extend the $|u_i\rangle$ to a orthonormal basis such that the matrix $U$ whose columns are the basis states has at most $k$ eigenvalues different from 1. This allows us to write $U = V(D \oplus I_{2^n-k})V^*$, where $D = D(e^{iw_0}, \ldots, e^{iw_{k-1}})$. Let $V_i$ be unitary operations representable by $b(g, n; 1)$ 2-qubit gates which take $|0\rangle$ to the $i$'th column of $V$ (the $i$'th eigenvector of $U$). Then

$$U = V_k I_{w_k} V_k^* \ldots V_1 I_{w_1} V_1^*.$$

∎

To obtain a general upper bound on $b(g, n; k)$ still requires bounding $b(g, n; 1)$. We show that $b(2, n; 1) = O(n2^n)$, thus improving the upper bound given in [1] by a factor of $n^2$ for the case $k = 2^n$.

**Theorem 3.3** $b(2, n; 1) = O(n2^n)$.

**Proof.** Suppose we would like to map $|0\rangle$ to $|u\rangle$. Write

$$|u\rangle = \sum_{b \in 2^{[n-1]}} |b\rangle(a(b, 0)|0\rangle + a(b, 1)|1\rangle).$$

Let $A^2(b) = |a(b, 0)|^2 + |a(b, 1)|^2$. Then $\sum_b A^2(b) = 1$. The idea is as follows: Using induction, apply an operation to the first $n - 1$ bits, making sure that the amplitude of $|b\rangle$ is $A(b)$. Next apply a conditional operation on the last bit to distribute the amplitude correctly for each $|b\rangle$.

More formally, let $|v\rangle = \sum_{b \in 2^{[n-1]}} A(b)|b\rangle$. Choose a unitary operator $U$ on $n - 1$ qubits which maps $|0\rangle$ to $|v\rangle|0\rangle$ using at most $b(2, n-1; 1)$ gates. Next choose conditional operations $U_b$ on $n$ qubits which have the property that $U_b|b'\rangle = |b'\rangle$ unless the first $n - 1$ bits of $b'$ are $b$. In that case define

$$U_b|b0\rangle = \frac{1}{A(b)}|b\rangle(a(b, 0)|0\rangle + a(b, 1)|1\rangle).$$

The composition of $U$ with each of the $U_b$ has the desired effect on $|0\rangle$. It remains to determine the complexity of $U_b$. Note that we have made no requirement on $U_b|b1\rangle$. This means that without loss of generality, $\det(U_b) = 1$. This implies that $U_b$ is one of the conditional operators $\wedge_{n-1}(V_b)$ with $V_b$ in $SU(2)$ discussed in [1]. It is shown there that $\wedge_{n-1}(V_b)$ can be constructed with $O(n)$ gates. This gives
$$b(2,n;1) \leq cn2^{n-1} + b(2, n-1; 1),$$
for some constant $c$. Thus $b(2,n;1) = O(n2^n)$. ∎

Lower bounds on $b(2,n;k)$ can be obtained by using dimensional arguments. We repeat the argument given in [1], with the relevant details filled in to obtain a more general result.

**Theorem 3.4** $b(2,n;k) \geq \frac{k}{9}(2^{n+1} - k) - \frac{1}{3}n - \frac{1}{9}$.

**Proof.** Consider a quantum circuit implementing operator $U$ with the gates $g_i$ labeled by the corresponding unitary operators $U_i \in U_2$ acting on two qubits. Two circuits are structurally the same if the only difference is in the associated unitary operators. Note that there are finitely many structurally distinct circuits involving $b$ gates (at most $\binom{n}{2}^b$). The set of sequences $\langle\, U|0\rangle, \ldots, U|k-1\rangle\,\rangle$ achievable by a fixed structure $S$ is the range of a smooth function $f_S$ with domain the sequence $\langle\, \ldots, U_i, \ldots\,\rangle$. The range of $f_S$ is to be compared to $\mathcal{U}_k$ whose dimension is $k(2^{n+1} - k)$. Clearly, the dimension of the domain manifold is $b\dim(U_2)$. The dimension of $U_2$ is 16. The task is to determine an upper bound on the dimension of the range of $f_S$. The following lemmas reduce the initial estimate of $16b$.

**Lemma 3.5** *Any gate other than the first one adds at most* 15 *dimensions to the range of $S$.*

**Proof.** We can remove the global phase from $U_i$ for $i \geq 2$ without changing the range. In other words, except for $U_0$, we can restrict the $U_i$ to unitary operators with at least one fixed point. ∎

**Lemma 3.6** *If gate $g_i$ satisfies that one of its inputs is an output of another gate, then the effective contribution of $U_i$ to the dimension of the range of $S$ is at most* 12.

**Proof.** Without loss of generality assume that the input which comes from another gate is the *first* input of gate $g_i$. For every $V \in U_2$, there is a neighborhood of $V$ that can be parametrized in the form $XW$, where $W$ is an operator which acts only on the first input qubit and $X$ lives in a submanifold of dimension 12. This follows from a fundamental theorem for homogeneous manifolds, see [11], page 120. The operator $W$ acts only on one output of the previous gate and can therefore be absorbed by that gate. In the complete circuit, one can work one's way from the output to the input, restricting $U_i$ to one of a finite number of local neighborhoods covering $U_2$, restricting $f_S$ to the corresponding $X$'s and moving the $W$'s to the previous gate. The result is a decomposition of the domain of $f_S$ to a finite union of restrictions to smaller dimensional spaces without affecting the union of the ranges. ∎

**Lemma 3.7** *If gate $g_i$ satisfies that both of its inputs are outputs of other gates, then the effective contribution of $U_i$ to the dimension of the range of $S$ is at most 9.*

**Proof.** The argument is the same as that for Lemma 3.6, except that the group acting independently on each input bit is used. This group has dimension 7, so there are still 9 dimensions to parametrize the remainder of $U_{g_i}$ in local neighborhood patches. ∎

The dimension of the modified domain of $f_S$ can now be counted as follows: All gates contribute 9 dimensions for a total of $9b$ dimensions. At most $n$ inputs do not come from a previous gate. Those $n$ inputs effectively contribute $3n + 1$ dimensions, where the $+1$ comes from the phase of the first gate.

The arguments given so far show that with $b$ gates, unless $9b + 3n + 1 \geq d$, where $d$ is the dimension of the range manifold, the measure of the range of $f_S$, being a finite union of submanifolds of dimension at most $9b + 3n + 1$, is 0. This implies that $b(2, n; k) \geq \frac{k}{9}(2^{n+1} - k) - \frac{1}{3}n - \frac{1}{9}$. ∎

## 4 Approximation of unitary operations by quantum circuits

We begin by considering approximability by quantum circuits of unitary operators in the $||\cdot||_{2,k}$ norm. Let $U_{(n,k,g,b,\epsilon)}$ be the set of unitary operators $U$ such that there exists a circuit in $G_g^{(b,n)}$ which computes a unitary operator $V$ with $||U - V||_{2,k} \leq \epsilon$. Let $\text{Exp}(2, x) = 2^x$ and $\rho(x) = x\sqrt{1 - x^2/4}$.

**Theorem 4.1** *For $\alpha > 0$ and $(1 + \alpha)\epsilon < \sqrt{2}$,*

$$\mu_k(U_{n,k,g,b,\epsilon}) \leq \text{Exp}\left(2, 2^{4g}b(\log(b) + \log(n) + \log(2/(\alpha\epsilon)))\right)$$

$$- k \left(2^n \log(1/\rho((1+\alpha)\epsilon) - \log(1/(1-(1+\alpha)^2\epsilon^2/2)))\right).$$

The constants in the bound of Theorem 4.1 can be improved by more careful estimation. The main consequence of the theorem is the following corollary.

**Corollary 4.2** *For fixed $g$ and $\epsilon < \sqrt{2}$, If $b \log(b) = o(k 2^n)$ then $\mu_k(U_{n,k,g,b,\epsilon}) = o(1)$.*

Note that $\sqrt{2}$ is the expected distance of a random state to a fixed state, so that if $b \log(b) = o(k 2^n)$, then for most $|u\rangle$ the circuits with less than $b$ gates can do essentially no better at mapping $|0\rangle$ to $|u\rangle$ than a random unitary operator.

**Proof of Theorem 4.1.** The proof is based on a covering argument. We first replace the unitary operators implemented by a single gate by a finite set of operators which is sufficiently dense. The number of distinct circuits is now finite. With the right choice of parameters, the volume of unitary operators that can be approximated is then given by the sum of the volume of suitable balls around each of the finite set of operators which can be represented exactly.

We first compute the volume of a $\delta$-ball for the $||\ ||_{2,k}$ norm relative to $\mu_k$.

**Lemma 4.3** *Let $U$ be a unitary operator and $B(U, \delta, k) = \{V \in U_n \mid ||V - U||_{2,k} < \delta\}$. Then*

$$\begin{aligned}
\mu_k(B(U,\delta,k)) &\leq (1-\delta^2/2)^{-k} \prod_{i=0}^{k-1} \frac{\left(\delta\sqrt{1-\delta^2/4}\right)^{2^{n+1}-1-2i}}{\sqrt{2^{n+1}-1-2i}} \\
&\leq (1-\delta^2/2)^{-k} \left(\delta\sqrt{1-\delta^2/4}\right)^{k 2^n}.
\end{aligned}$$

**Proof.** For a logical formula $\phi$, let $[\phi] = 1$ if $\phi$ is true, and $[\phi] = 0$ otherwise. Let $|u_0\rangle, \ldots, |u_{k-1}\rangle$ be the first $k$ columns of $U$. We have

$$\begin{aligned}
\mu_k(B(U,\delta,k)) &= \int d\mu_k(\mathbf{u})[\mathbf{u} \in \phi_k(B(U,\delta,k))] \\
&\leq \int d\mu_{k-1}(\mathbf{v})[\mathbf{v} \in \phi_{k-1}(B(U,\delta,k-1))] \\
&\quad \int d\mu_1(|u\rangle)[\ |u\rangle \perp \mathbf{v} \text{ and } ||u\rangle - |u_{k-1}\rangle| < \delta]
\end{aligned}$$

$$\leq \int d\mu_{k-1}(\mathbf{v})[\mathbf{v} \in \phi_{k-1}(B(U, \delta, k-1))]$$

$$(1 - \delta^2/2)^{-1} \frac{\left(\delta\sqrt{1 - \delta^2/4}\right)^{2^{n+1} - 2k + 1}}{\sqrt{2^{n+1} - 2k + 1}},$$

where we have used Lemma 2.4. The first inequality now follows by induction. The second one follows by observing that $k \leq 2^n$, the dimension of the space. ∎

**Lemma 4.4** *For $\delta > 0$, there is a subset $U_{n,\delta}$ of $U_n$ with*

$$\left|U_{n,\delta}\right| = m(\delta, n) \leq \left(\frac{2}{\delta}\right)^{2^{4n}}$$

*such that for any $U \in U_n$, there exists $V \in U_{n,\delta}$ with $||U - V||_2 \leq \delta$.*

**Proof.** For $U \in U_n$, each entry $U_{i,j}$ has norm at most 1. We can construct a set $\mathcal{A}$ of operators (not necessarily unitary) by selecting each of the $2^n$ matrix entries from among $\frac{1}{2\rho^2}$ equally spaced values in the (complex) unit square. The cardinality of $\mathcal{A}$ is $\left(\frac{1}{2\rho^2}\right)^{2^{2n}}$. For any $U$ there exists a $B \in \mathcal{A}$ such that $U - B$'s entries are within $\rho$ of 0. Hence $||U - B||_2 \leq ||U - B||_F \leq \rho 2^n$. Choose $\rho 2^n = \delta/2$. Construct $U_{n,\delta}$ by selecting for each $A \in \mathcal{A}$ a nearest unitary matrix (according to the $|| \ ||_F$ norm). This increases the maximum distance by at most a factor of 2, so that our choice of $\rho$ is good. The cardinality of the resulting set is less than the stated bound. ∎

Choose $\delta = \frac{\alpha\epsilon}{b}$ in Lemma 4.4. If $U \in U_{n,k,g,b,\epsilon}$, then by Lemma 4.4 and 2.3, there exists $V \in G_{g,\delta}^{(b,n)}$ such that $||V - U||_{2,k} \leq (1 + \alpha)\epsilon$. Here we used $G_{g,\delta}$ to denote the set of operations achievable by a single gate whose unitary operator is in $U_{g,\delta}$. The volume of $U_{n,k,g,b,\epsilon}$ is therefore at most the sum of the volumes of the $(1 + \alpha)\epsilon$ balls (relative to $||\cdot||_{2,k}$) around each member of $G_{g,\delta}^{(b,n)}$. The relative volume of these balls is bounded in Lemma 4.3. Since $\left|G_{g,\delta}^{(b,n)}\right| \leq (m(\delta, g)\binom{n}{g})^b$ we get

$$\begin{aligned}\mu_k(U_{n,k,g,b,\epsilon}) &\leq \left(m(\delta, g)\binom{n}{g}\right)^b (1 - (1+\alpha)^2\epsilon^2/2)^{-k} \rho((1+\alpha)\epsilon)^{k2^n} \\ &\leq \operatorname{Exp}\Big(2, b(2^{4g}(\log(b) + log(2/(\alpha\epsilon))) + \log(n)) \\ &\quad - k\left(2^n \log(1/\rho((1+\alpha)\epsilon) - \log(1/(1 - (1+\alpha)^2\epsilon^2/2))\right)\Big).\end{aligned}$$

From the point of view of quantum computation, a bound on approximability of density matrices on a subset of the bits is critical. Let $U_{n,l,k,g,b,\epsilon}$ be the set of unitary operators $U$ such that there exists a circuit in $G_g^{(b,n)}$ which computes a unitary operator $V$ with $v_{l,k}(U,V) \leq \epsilon$.

**Theorem 4.5** *For $\alpha > 0$ and $2(1+\alpha)\epsilon < 1$,*

$$\begin{aligned}\nu_{l,k}(U_{n,l,k,g,b,\epsilon}) &\leq \text{Exp}\left(2, b(2^{4g}(\log(b) + log(4/(\alpha\epsilon))) + \log(n))\right. \\ &\left. - k2^{l-1}\left(\log(1/2(1+\alpha)\epsilon))\right)\right).\end{aligned}$$

Asymptotically, Theorem 4.5 has the following consequence:

**Corollary 4.6** *For fixed $g$ and $\epsilon < \frac{1}{2}$, If $b\log(b) = o(k2^l)$ then $\mu_{l,k}U_{n,l,k,g,b,\epsilon} = o(1)$.*

Corollary 4.6 is not tight in the same way that Corollary 4.2 is. The minimum average total variation distance to a distribution on $N$ points is $\frac{2}{e} > \frac{1}{2}$. This gap can be improved by more careful estimates in Lemma 2.5.

**Proof of Theorem 4.5.** The proof follows the same outline as the proof of Theorem 4.1. The analog of Lemma 4.3 is as follows:

**Lemma 4.7** *Let $U$ be a unitary operator and $B(U, \delta, l, k) = \{V \in U_n \mid v_{l,k}(V, U) < \delta\}$. Then*

$$\nu_{l,k}(B(U, \delta, l, k)) \leq (2\epsilon)^{k(2^l-1)}.$$

By Lemma 2.1, we can also use Lemma 4.4 for the $v_{l,k}$ norm, taking care to compensate for the factor of 2 in the inequality of Lemma 2.1. Choose $\delta = \alpha\epsilon/2b$ in Lemma 4.4. The same argument used in Theorem 4.1 gives

$$\begin{aligned}\nu_{l,k}U_{n,l,k,g,b,\epsilon} &\leq \left(m(\delta, g)\binom{n}{g}\right)^b (2(1+\alpha)\epsilon)^{k(2^l-1)} \\ &\leq \text{Exp}\left(2, b(2^{4g}(\log(b) + \log(4/(\alpha\epsilon))) + \log(n))\right. \\ &\left. - k2^{l-1}\left(\log(1/2(1+\alpha)\epsilon))\right)\right).\end{aligned}$$

∎

## 5 Classical approximation problems

There are two classical approximation problems to be considered.

**Definition.** Let $f : D \subseteq 2^{[n]} \to \{0,1\}$ be a boolean decision problem and $g : D \to 0,1$ a probabilistic function. Then $g$ *(probabilistically) decides $f$ with advantage $q$* if for all $b \in D$

$$\operatorname{Prob}(g(b) = f(b)) \geq \frac{1}{2}\left(1 + \frac{1}{q}\right).$$

If $g$ decides $f$ with advantage $q$, then $f(b)$ can be obtained correctly with high probability by $O(q^2)$ independent evaluations of $g$.

To probabilistically decide $f$ with advantage $q$ by a quantum circuit (that is by a unitary operator $U \in U_{n'}$ with $n' \geq n$) means that for $b \in D$, a measurement of the first bit of $U|b,0\rangle$ yields $f(b)$ with probability at least $\frac{1}{2}(1 + \frac{1}{q})$.

**Definition.** Let $f : D \subseteq 2^{[n]} \to 2^{[n]}$ be a function and $g : D \to 2^{[n]}$ a probabilistic function. Then $g$ *guesses $f$ with advantage $q$* if for all $b \in D$,

$$\operatorname{Prob}(g(b) = f(b)) \geq \frac{1}{q}$$

A *checking* oracle for $f$ is a function $h$ which on input $b, b'$ decides whether $f(b) = b'$. If $g$ guesses $f$ with advantage $q$ and we have a checking oracle for $f$, then we can obtain the values of $f$ with high probability by $O(q)$ independent evaluations of $g$, checking each evaluation using the oracle.

To guess $f$ with advantage $q$ by a operator $U \in U_{n'}$ with $n' \geq n$ means that $\sum_{b'} |\langle f(b)b'|U|b0\rangle|^2 \geq \frac{1}{q}$. Note that the sum is the square amplitude of the projection of the result onto the space determined by setting the first $n$ bits to $f(b)$.

We now show that there are decision and guessing problems which are difficult for quantum computation. To avoid trivial cases, we assume that $g \geq 2$ and $b \geq 2$. We can also assume $bg \geq n' - n$. Otherwise some of the additional input qubits do not appear as an input to a gate and therefore do not contribute to the computation.

**Theorem 5.1** *The fraction of decision problems with domain $D$ decided by elements of $G_g^{(b,n')}$ with advantage $q$ is at most*

$$\operatorname{Exp}(2, (\log(b) + \log(q) + \log(n))b2^{4g+1} - |D|).$$

**Proof.** There are $2^{|D|}$ decision problems for domain $D$. Consider a fixed $U \in U_{n'}$. How many decision problems can be decided probabilistically by $V \in U_n$ with $||V - U||_2 < \delta$ with advantage $q$? Let $f$ be one decision problem with domain $D$ solved by a $V \in U_{n'}$ with $||V - U||_2 < \delta$. For $||W - U||_2 < \delta$ and $b \in D$ we have

$$\begin{aligned} v_l(W|b\rangle - V|b\rangle) &\leq 2||W|b\rangle - V|b\rangle|| \\ &= 2|(W - U)|b\rangle + (U - V)|b\rangle| \\ &\leq 2|(W - U)|b\rangle| + 2|(U - V)|b\rangle| \\ &\leq 4\delta. \end{aligned}$$

Thus if $\delta < \frac{1}{q}$, then $W$ either does not decide any decision problem with domain $D$ with advantage $q$ or it decides $f$ with advantage $q$. This implies that the number of decision problems that can be decided by members of $G_g^{(b,n')}$ is at most the number of $\frac{1}{q}$ balls required to cover $G_g^{(b,n')}$. Using Lemmas 2.3 and 4.4 with $\delta = \frac{1}{bq}$, this number is at most

$$\begin{aligned} m\left(\frac{1}{bq}, g\right)^b \binom{n'}{g}^b &\leq (2bq)^{b2^{4g}} n'^b \\ &\leq \mathrm{Exp}(2, (1 + \log(b) + \log(q) + \log(n'))b2^{4g}) \\ &\leq \mathrm{Exp}(2, (\log(b) + \log(q) + \log(n))b2^{4g+1}), \end{aligned}$$

where we have used the assumptions rather loosely to eliminate $n'$ in favor of $n$. ∎

**Corollary 5.2** *For a sequence of domains $D_n \subseteq 2^{[n]}$, there are decision problems that cannot be decided by quantum circuits with advantage $q$ unless $(\log(b) + \log(q) + \log(n))b = \Omega(|D_n|)$.*

Note that for interesting domains, $D_n$ is exponential in $n$.

We can obtain a similar result for guessing functions.

**Theorem 5.3** *The fraction of functions $f : D \subseteq 2^{[n]} \to 2^{[n]}$ guessed with advantage $q$ by operators in $G_g^{(b,n')}$ is at most*

$$\mathrm{Exp}(2, (\log(b) + \log(q) + \log(n))b2^{4g+1} - (n - \log(4q))|D|).$$

**Proof.** The number of functions considered is $2^{n|D|}$. We obtain a bound on the number of functions guessed with advantage $q$ by operators $V$ within $\delta$ in 2-norm of a fixed operator $U \in U_{n'}$. For a vector $u$, let $P_b(u)$ be the square amplitude of the projection of $u$ onto the space spanned by the states $|b, b'\rangle$. We have $\sum_b P_b(u) = |u|^2$.

Let $||V - U||_2 < \delta$. Let $b \in D$, $|u\rangle = U|b\rangle$ and $|v\rangle = V|b\rangle$. Let

$$S(U, b) = \{c \mid P_c(|u\rangle) \geq 1/4q\},$$
$$S(V, b) = \{c \mid P_c(|v\rangle) \geq 1/q\}.$$

If $V$ guesses $f$ with advantage $q$, then $f(b) \in S(V, b)$ for each $b \in D$. We have $||u\rangle - |v\rangle| < \delta$. We would like to estimate $|S(V, b) \setminus S(U, b)|$.

$$||u\rangle - |v\rangle|^2 = \sum_c P_c(|u\rangle - |v\rangle)$$
$$\geq \sum_{c \in S(V,b) \setminus S(U,b)} P_c(|u\rangle - |v\rangle)$$
$$\geq \sum_{c \in S(V,b) \setminus S(U,b)} \frac{1}{4q}$$
$$= |S(v, b) \setminus S(U, b)| \frac{1}{4q},$$

where we used

$$P_c(|u\rangle - |v\rangle)^{1/2} \geq |P_c(|v\rangle)^{1/2} - P_c(|u\rangle)^{1/2}|$$
$$\geq \frac{1}{2\sqrt{q}}.$$

If we choose $\delta^2 < \frac{1}{4q}$, then $|S(V, b) \setminus S(U, b)| < 1$, so that $S(V, b) \subseteq S(U, b)$. We can estimate $S(U, b)$ by observing that

$$1 = |||u\rangle||^2$$
$$= \sum_c P_c(|u\rangle)$$
$$\geq \sum_{c \in S(U,b)} P_c(|u\rangle)$$
$$\geq \sum_{c \in S(U,b)} \frac{1}{4q}$$
$$= |S(U, b)| \frac{1}{4q},$$

which implies that $|S(U, b)| \leq 4q$. With this $\delta$, the set of functions $f$ with domain $D$ guessed by any $V$ with $||V - U||_2 < \frac{1}{2\sqrt{q}}$ satisfies $f(b) \in S(U, b)$ for each $b \in D$, so its cardinality is at most $(4q)^{|D|}$.

The number of functions $f : D \subseteq 2^{[n]} \to 2^{[n]}$ that are guessed with advantage $q$ by members of $G_g^{\prime(b,n')}$ can now be bounded by

$$(4q)^{|D|} m\left(\frac{1}{2\sqrt{q}b}, g\right)^b \binom{n'}{g}^b$$

$$\leq (4q)^{|D|}(4\sqrt{q}b)^b 2^{4g} n'^b$$

$$\leq \mathrm{Exp}(2, (2 + \log(b) + \frac{1}{2}\log(q) + \log(n'))b 2^{4g} + \log(4q)|D|)$$

$$\leq \mathrm{Exp}(2, (\log(b) + \log(q) + \log(n))b 2^{4g+1} + \log(4q)|D|).$$

∎

**Corollary 5.4** *For a sequence of domains $D_n \subseteq 2^{[n]}$, there are functions $f : D_n \subseteq 2^{[n]} \to 2^{[n]}$ not guessed with advantage $q$ by any member of $G_g^{\prime(b,n')}$ unless*

$$(\log(b) + \log(q) + \log(n))b = \Omega((n - \log(4q))|D_n|).$$

Again, the interesting cases are where $|D_n|$ is exponential in $n$.

## A  Appendix

**Proof of Lemma 2.4.** Assume first that $m$ is the dimension of the full space. We can give an explicit integral for $\mu(B_{u,\epsilon} \cap S)$:

$$\mu(B_{u,\epsilon} \cap S) = \int_0^{\sin(\phi)=\epsilon\sqrt{1-\epsilon^2/4}} d\phi \sin(\phi)^{2m-2} 2 \frac{\pi^{m-1/2}}{\Gamma(m-1/2)}$$

for $\epsilon \leq \sqrt{2}$. The integral is obtained by parametrizing the spheres $\{v \mid |v - u| = \delta \leq \epsilon, |v| = 1\}$ by the angle (in real Euclidean space) between the $v$ and $u$. We can write

$$\int_0^{\sin(\phi)=\delta} d\phi \sin(\phi)^{2m-2} = \int_0^\delta dx \frac{x^{2m-2}}{\sqrt{1-x^2}}$$

$$\leq \int_0^\delta dx \frac{x^{2m-2}}{\sqrt{1-\delta^2}}$$

$$\leq \frac{\delta^{2m-1}}{(2m-1)\sqrt{1-\delta^2}}.$$

Substituting the upper bound in terms of $\epsilon$ gives

$$\mu_1(B_{u,\epsilon} \cap S) \leq \frac{\Gamma(m)}{(2m-1)\sqrt{\pi}\Gamma(m-1/2)} \frac{\left(\epsilon\sqrt{1-\epsilon^2/4}\right)^{2m-1}}{(1-\epsilon^2/2)}.$$

Using the Sterling approximation we can estimate for $m \geq 2$

$$\sqrt{2\pi(m-1)} \left(\frac{m-1}{e}\right)^{m-1} \leq \Gamma(m) \leq \sqrt{4\pi(m-1)} \left(\frac{m-1}{e}\right)^{m-1}.$$

Thus

$$\frac{\Gamma(m)}{\Gamma(m-1/2)} \leq \sqrt{2(m-1)/e} \left(\frac{m-1}{m-3/2}\right)^{m-1}$$

$$= \sqrt{2(m-1)/e} \left(1 - \frac{1}{2(m-1)}\right)^{-(m-1)}.$$

To obtain good upper bounds, use the inequalities $1/(1-x) \leq e^x/(1-x^2)$ and $(1-x)^m \geq (1-mx)$.

$$\frac{\Gamma(m)}{\Gamma(m-1/2)} \leq \sqrt{2(m-1)} \frac{1}{1 - 1/(4(m-1))}.$$

Therefore

$$\mu_1(B_{u,\epsilon} \cap S) \leq \frac{\left(\epsilon\sqrt{1-\epsilon^2/4}\right)^{2m-1}}{\sqrt{(2m-1)\pi}(1-1/(4(m-1)))(1-\epsilon^2/2)}$$

$$\leq \frac{\left(\epsilon\sqrt{1-\epsilon^2/4}\right)^{2m-1}}{\sqrt{2m-1}(1-\epsilon^2/2)}$$

for $m \geq 3$ and $\epsilon < \sqrt{2}$.

To complete the proof of the lemma, we show that in the general case, $B_{u,\epsilon} \cap S \subseteq B_{v/|v|,\epsilon} \cap S$, where $v$ is the real projection of $u$ onto the subspace generated by $S$. We can assume that $v$ is non-zero, as otherwise for $\epsilon < \sqrt{2}$, $B_{u,\epsilon} \cap S = \emptyset$. Let $\cdot$ denote the real inner product. Let $x \in B_{u,\epsilon} \cap S$. Then $u \cdot x = v \cdot x \leq \frac{v}{|v|} \cdot x$. Note that $u$, $x$ and $v$ are in the same halfspaces determined by the hyperplane perpendicular to $v$ and the one perpendicular to $x$ (otherwise $|u-x| > \sqrt{2}$). Let $\phi$ and $\psi$ be the angles between $u$ and $x$ and between $v$ and $x$, respectively. The inequality between the inner products means that $\phi \geq \psi$. Since $u$, $x$ and $v$ are unit vectors, $|u-x| \geq |v-x|$, which gives the result. ∎

**Proof of Lemma 2.5.** Let $v \in \Delta(N)$. Let $H$ be the hyperplane which contains $\Delta(N)$. Let $B$ be the $\epsilon$ ball in the $L_1$ metric around $v$. To compute $\mu(B \cap H)$ we can shift $v$ to the origin and consider each orthant separately. After projection onto all but one of the negative coordinates, the intersection of $B \cap H$ with an orthant with $k$ positive coordinates is equivalent to the set

$$\left\{ x \mid x_i \geq 0, \sum_{i=1}^{k} x_i \leq \epsilon/2, \sum_{i=k+1}^{N-1} x_i \leq \sum_{i=1}^{k} x_i \right\}.$$

The total projected volume of the orthants is therfore given by

$$\begin{aligned}
\mu(B \cap H) &= \sum_k \binom{N}{k} \int_0^{\epsilon/2} dt \frac{\mu(\Delta(k))}{\sqrt{k}} t^{k-1} \frac{1}{\Gamma(N-k)} t^{N-k-1} \\
&= \sum_k \binom{N}{k} \frac{1}{(N-1)\Gamma(k)\Gamma(N-k)} (\epsilon/2)^{N-1} \\
&= \frac{1}{\Gamma(N)} \sum_k \binom{N}{k} \binom{N-2}{k-1} (\epsilon/2)^{N-1} \\
&\leq \frac{1}{\Gamma(N)} (2\epsilon)^{N-1}.
\end{aligned}$$

Since the projected volume of $\Delta(N)$ is $\frac{1}{\Gamma(N)}$, the result follows. ∎

**Lemma A.1** *Let $X$ and $Y$ be $n \times m$ operators such that $X^*X = Y^*Y$. Then there exists a unitary operator $U$ such that $UX = Y$.*

**Proof.** Let $Y = WDV$ be the singular value decomposition of $Y$, with $D$ a non-negative diagonal matrix. By a change of coordinates, we can assume that $V = I$. This gives $Y^*Y = D^*D$. $X^*X = D^*D$ implies that the columns of $X$ are orthogonal, of lengths corresponding to the entries of $D$. Hence there is a unitary transformation $W'$ such that $W'X = D$. Letting $U = WW'$ gives the lemma. ∎

**Acknowledgements.** This work benefited greatly from many helpful discussions with Prasad Chalasani, Richard Hughes, Raymond Laflamme, Madhav Marathe, and Clint Scovel.